\documentclass[fleqn,twoside]{article}
\usepackage{espcrc2}
\usepackage{amssymb}
\usepackage{graphicx}

\newcommand{\AmS}{{\protect\the\textfont2
  A\kern-.1667em\lower.5ex\hbox{M}\kern-.125emS}}
\title{ Screening Masses in Dimensionally
          Reduced (2+1)D Gauge Theory }

\begin{document}
\author{
P. Bialas, \address{Inst. of Comp. Science, Jagellonian University\\
33-072 Krakow, Poland}
A. Morel, \address{Service de Physique Th\'eorique de Saclay, CE-Saclay,\\
F-91191 Gif-sur-Yvette Cedex, France}
B.Petersson, 
\address[BI]{Fakult\"at f\"ur Physik, Universit\"at Bielefel\\
P.O.Box 100131, D-33501 Bielefeld, Germany} 
 and K. Petrov\addressmark[BI] \thanks{The contribution was presented by K.P.} 
}
\begin{abstract} 
We discuss the screening masses and residue factorisation of the $SU(3)$ $(2+1)D$ theory in the dimensional reduction formalism. The phase structure of the reduced model is also investigated.

\end{abstract}
\maketitle
\section{Introduction } 
We present a numerical study of a 2D SU(3) gauge theory with adjoint scalar 
fields, defined by dimensional reduction of pure gauge QCD in (2+1)D at high 
temperature. We show that the reduced model not only reproduces Polyakov
 Loops and their correlations but also gives a good approximation for the
 spatial string tension \cite{paper1}.  We also show that the correlations 
between Polyakov loops are
saturated by two colourless bound states. Their 
contributions in correlation functions of
local composite operators $\phi_n$ respectively of degree $n=2p$ and $2p+1$ in the scalar fields
($p=1,2$) fulfill factorization. The contributions of two particle states 
are detected. Their size agrees with the estimates based on a 
meanfield-like decomposition of the $p=2$ operators into polynomials in
$p=1$ operators. In contrast to the naive picture of Debye screening, 
no sizable signal in any $\phi_n$ correlation can be attributed
to $1/n$ times a Debye screening length associated with $n$
elementary fields. These results are quantitatively consistent with the 
picture of scalar ``matter'' fields confined within colourless boundstates 
whose residual ``strong'' interactions are very weak \cite{paper2}.
\section{Dimensional Reduction and its Validity Region}

\begin{figure}
\includegraphics[width=70mm,height=70mm]{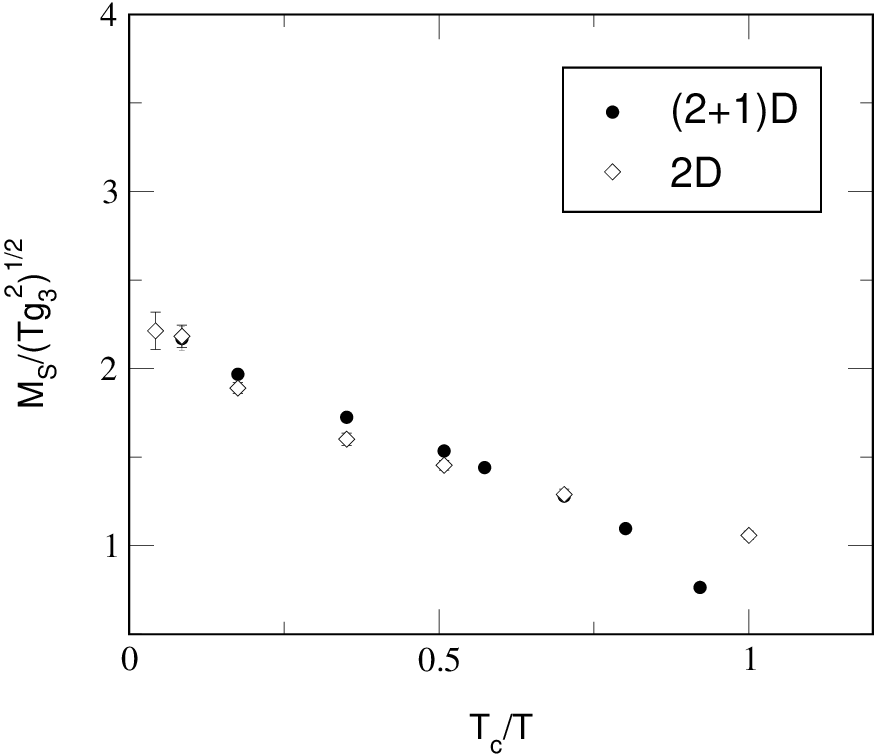}
\caption{$M_S$ is a mass in physical units}
\label{masses}
\end{figure}

The reduced lattice action $S_{eff}$ results from integrating out the infrared 
convergent non-static modes
of the gauge fields of the $(2+1)D$ theory at temperature $T$ and coupling
$g_3$. $S_{eff}$  depends on $2D$ gauge fields and on a ``Higgs''
field (noted $\phi(x)$), the static components respectively of the spatial
and time $3D$ gauge fields.
In the large $T$ limit and for momenta $p\ll T$ (large distance physics), 
$S_{eff}$ can be written (see \cite{paper1} for
details)
$$
S_{eff}=S_U+S_{U,\phi}+V(\phi).
$$
Here $S_U$ is pure gauge, $S_{U,\phi}$ is the gauge invariant $\phi$
kinetic term, and $V(\phi)$ is a local potential involving the  quadratic and 
quartic self-couplings $h_2, h_4$, which are computed in \cite {paper1} as functions of $g_3$ anf $T$ or, equivalently - of the  
$(2+1)D$ lattice parameters: coupling $\beta_3$, time-direction extension $L_0$ and spacing $a$ 
$$
\beta_3 = \frac{6}{a g^2_3} ; \,\,\, L_0 = \frac{1}{aT};
\,\,\, \tau \equiv \frac{T}{g^2_3} = \frac{\beta_3}{6 L_0}.
$$
To determine the region of validity of the dimensional reduction, we measured 
correlations of the Polyakov Loops, which are static operators 
\[ 
L (\bar x) =  \frac{1}{3}Tr \, exp({i\phi(\bar x) / \sqrt\tau}),
\]
and the string tension, extracted from the spatial Wilson Loops.
We performed the $2D$ simulations  
on a $32^2$ lattice with $L_0=4$, using 
Multi-hit Metropolis as well as Hybrid Metropolis-Heatbath 
algorithms. The method turned out to be valid for temperatures 
down to $T=1.5T_c$, $T_c$ being the deconfinement temperature of the $(2+1)D$ model \cite{lego} (Fig.\ref{masses})
In
order to make sure that we are close to the scaling region, we also performed 
some measurements with a lattice spacing twice smaller. That 
justifies keeping $L_0=4$ and comparing our $2D$ data with  
the $(2+1)D$ data of \cite{lego}. 
\section{General properties of the reduced model}
\begin{figure}
\includegraphics[width=70mm,height=70mm]{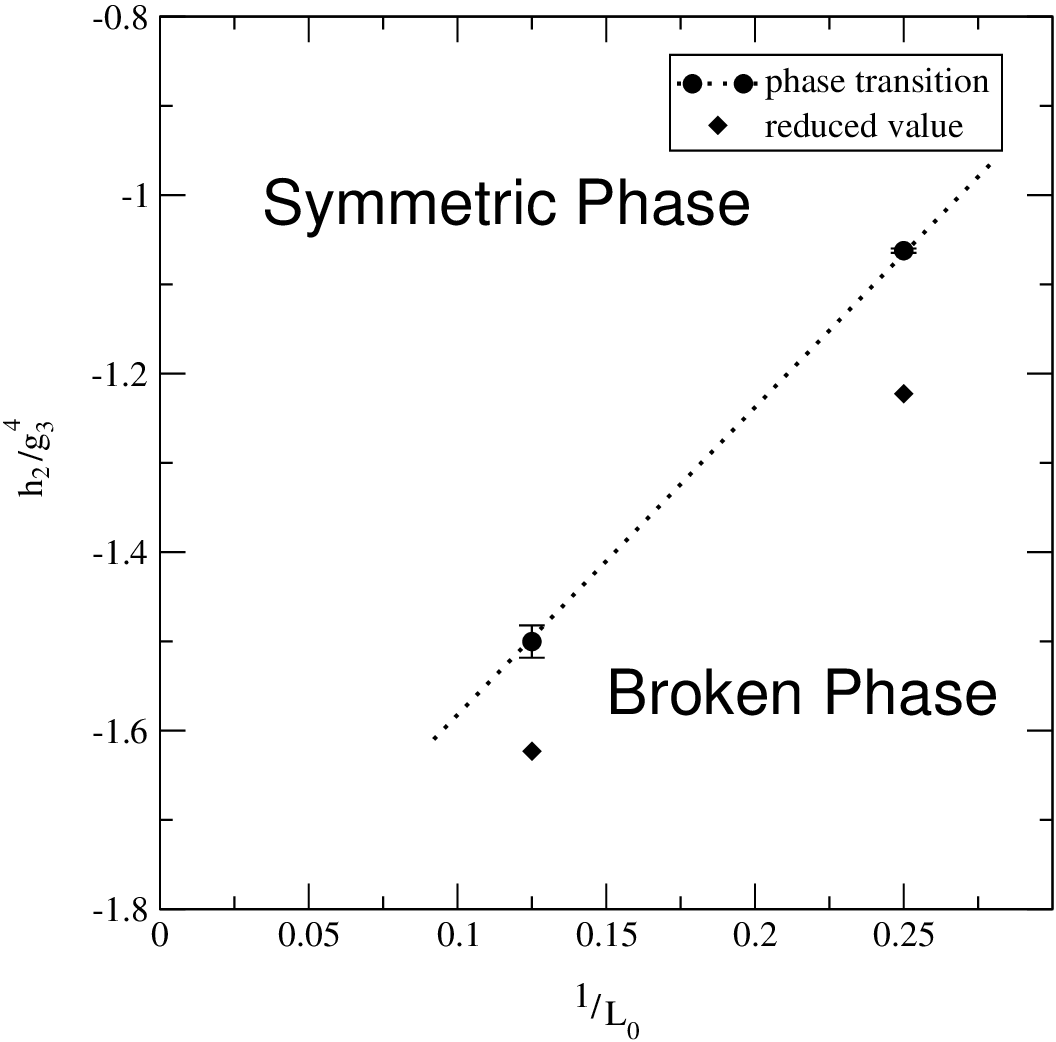}             
\caption{Phase diagram of the model. The line drawn through the measured transition points is just 
to visualize the phases locations.}
\label{phase}
\end{figure}

The system has a strong first order phase transition 
with respect to $R_{\tau}$, the imaginary time inversion symmetry of the 
original model, i.e. the $\phi \to -\phi$ symmetry, for which $tr\phi^3$ is
an order parameter. We find that in the thermodynamical limit the 
parameters of the reduced model correspond to a 
point in the $\it broken$ $R_{\tau}$ phase 
(but very close to the transition line). 
To avoid getting into the broken phase we start with weak field configurations, and check that the system stays in
the (metastable) phase monitored by $<tr\phi^3>=0$. The phase diagram in
the $h_2,aT$ plane is sketched in Fig.\ref{phase} : at two temperatures the
value of the quadratic self coupling $h_2$ at the transition is
 compared with that for the reduced model, and the phases are indicated.

\begin{figure}
\includegraphics[width=70mm, height=70mm]{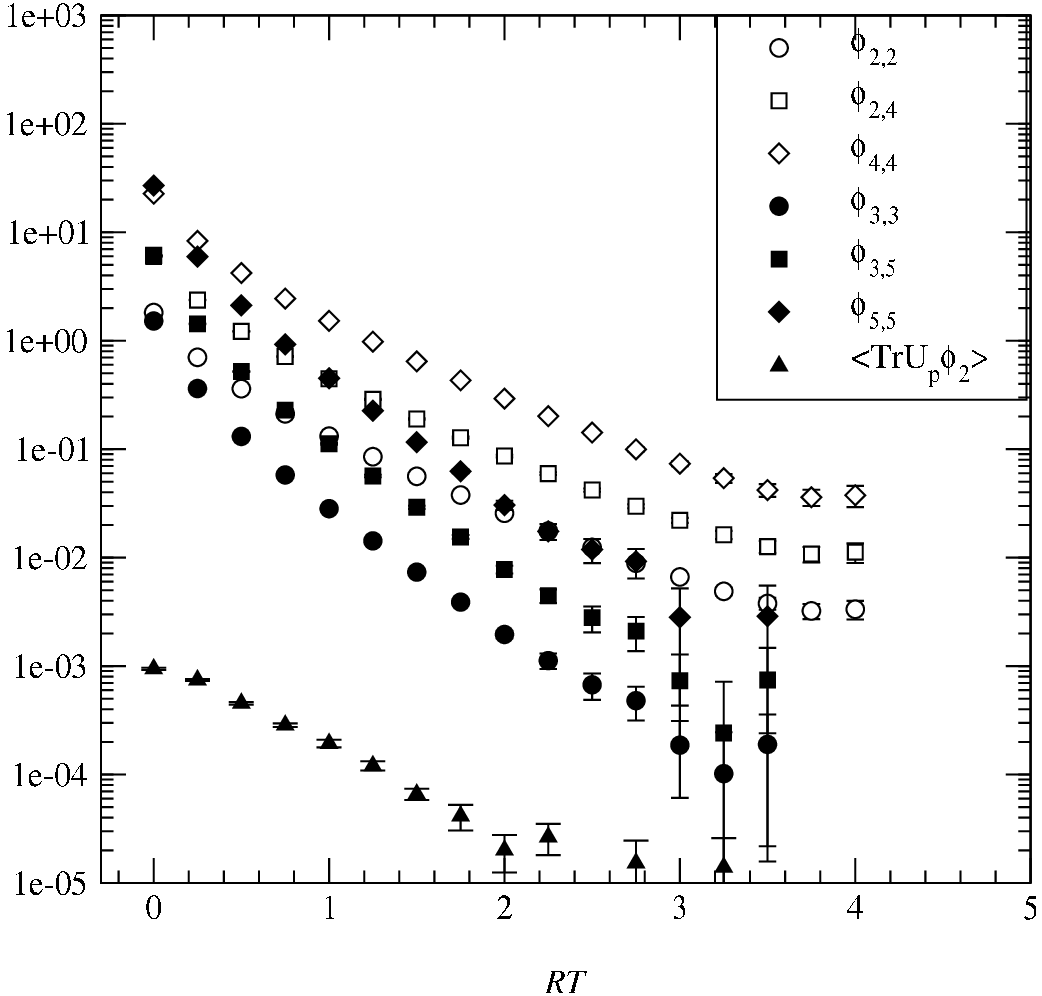}
\caption{}
\label{corrfig}
\end{figure}

 Careful study of the correlations 
\[
\phi_{n,m} \equiv \langle Tr\phi^n(x) Tr\phi^m(y)\rangle
\]
reveals two channels called $S$ and $P$ (Fig.\ref{corrfig}), associated with operators
respectively even and odd under $R_{\tau}$ i.e. in $\phi$. The mass $M_S$
measured in the even channel agrees with that measured from Polyakov
loops correlations (Fig. 1), in which the lowest power in $\phi$, namely
2, dominates.
The ratio of the corresponding screening masses does not show 
any clear tendency,
 being 1.8, 2.0, 1.7 and 1.6 in order of increasing temperature. It may still
 go to the value of 1.5 
at a very high temperature as expected from a naive composite gluon picture. 

One may also study the operators which change/do not change the sign under
 the other symmety operations, under which the action is invariant, 
e.g. reflection along one of the axis. Such operators however 
include both gauge and higgs operators. As it may be seen from 
the Fig.\ref{corrfig} the correlation 
between the plaquette and the $Tr\phi^2$ (triangles) is several orders of magnitude smaller than the $\phi_{2,2}$. 
Thus such measurements require significantly higher 
computational power and are in progress. 

\section{ Free Particles Model}

By comparing the size of the three different correlations, 
belonging to $S$ and $P$ channels we were able to show that 
the residue factorisation holds, as expected on general 
grounds when one particle propagates between two different 
states. This may be demonstrated by the following quantity:

$$
X_n \equiv \frac{\phi _{n,n} (r) \, \phi _{n+2, n+2} (r)}{\phi^2_{n, n+2} (r)}.
$$

\begin{figure}
\includegraphics[width=70mm, height=70mm]{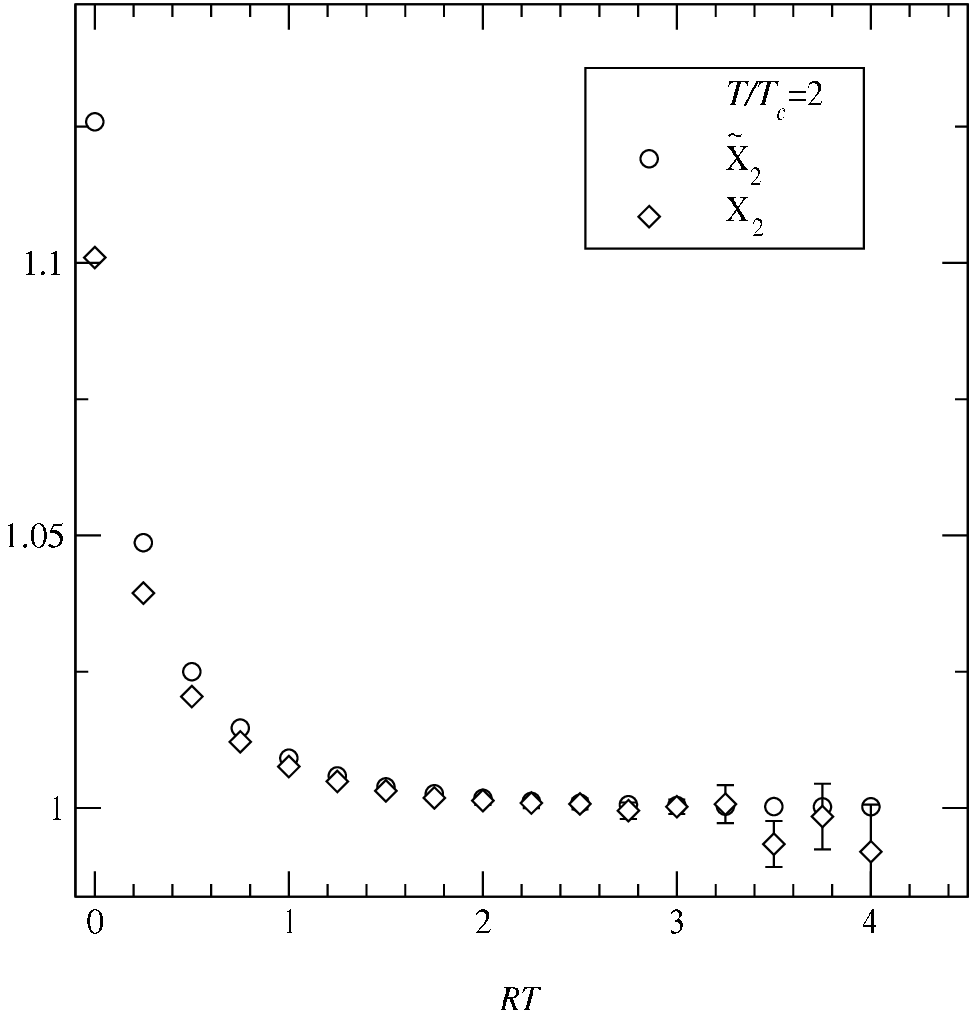}
\caption{Residue factorisation}
\label{X2}
\end{figure}

It should go to one at large $r$. Deviations at shorter distances are to a large extent compatible 
with the propagation of two particles, namely two $S$ or $S$ and 
$P$ respectively in $S$ and $P$ channels. A Wick-like treatment 
in the mean field approximation leads to, for example, the 
following result for the  $\phi_{4 , 4}$ 
$$
\phi_{4 , 4} (r) = \left( \frac{5}{4} \right)^2 \left[ \phi^2_2 \, \phi_{2 ,2} (r)+ \frac{1}{2} \phi^2_{2,2} (r) \right]
$$
The measured values of $X_2$ and those  $\tilde X_2$ which include the two $S$ correction to $\phi_{4,4}$ are successfully compared in Fig.\ref{X2}.


\begin{thebibliography} {2}

\bibitem{paper1}
P. Bialas, A. Morel, B. Petersson, K. Petrov and T. Reisz, ``High
Temperature 3D QCD: Dimensional Reduction at Work'', {\it Nucl.\ Phys.}
{\bf B581} (2000) 477.
\bibitem{paper2}                
P. Bialas, A. Morel, B. Petersson, K. Petrov and T. Reisz, ``QCD with Adjoint Scalars in 2D: Properties of the Colorless Scalar Sector'', {\it Nucl.\ Phys.}
{\bf B603} (2001) 369.
\bibitem{lego} C.Legeland, {\it PhD. Thesis},  ``Aspects of (2+1)
Dimensional Lattice Gauge Theory'' 
(University of Bielefeld, Germany, September 1998).
\end{thebibliography}
\end{document}